\newif\ifarxiv
\arxivtrue

\documentclass[conference]{IEEEtran}

\ifarxiv
\usepackage[frozencache,cachedir=.]{minted}
\else
\usepackage{minted}
\fi
\usepackage{cite}
\usepackage{amsmath,amssymb,amsfonts}
\usepackage{cleveref}
\usepackage{algorithmic}
\usepackage{graphicx}
\usepackage{textcomp}
\usepackage{xcolor}
\usepackage{xspace}
\usepackage{soul}
\renewcommand{\hl}[1]{\xspace}
\usepackage{url}

\newcommand\mdoubleplus{\ensuremath{\mathbin{+\mkern-10mu+}}}

\def\BibTeX{{\rm B\kern-.05em{\sc i\kern-.025em b}\kern-.08em
    T\kern-.1667em\lower.7ex\hbox{E}\kern-.125emX}}
\begin{document}

\title{
Scaling Program Synthesis Based \\
Technology Mapping
with Equality Saturation
}

\author{
    \IEEEauthorblockN{Gus Henry Smith, Colin Knizek, Daniel Petrisko \\
    Zachary Tatlock, Jonathan Balkind, Gilbert Louis Bernstein, Haobin Ni, Chandrakana Nandi}
    
    \IEEEauthorblockA{gussmith@cs.washington.edu, knizek@berkeley.edu, petrisko@cs.washington.edu, \\ ztatlock@cs.washington.edu jbalkind@ucsb.edu, gilbo@cs.washington.edu, hn42@cs.washington.edu, cnandi@cs.washington.edu}
}

\newcommand{\egraph}{\mbox{e-graph}\xspace}
\newcommand{\egraphs}{\mbox{e-graphs}\xspace}
\newcommand{\Egraph}{\mbox{E-graph}\xspace}
\newcommand{\Egraphs}{\mbox{E-graphs}\xspace}
\newcommand{\Eclass}{\mbox{E-class}\xspace}
\newcommand{\Eclasses}{\mbox{E-classes}\xspace}
\newcommand{\eclass}{\mbox{e-class}\xspace}
\newcommand{\eclasses}{\mbox{e-classes}\xspace}
\newcommand{\Enode}{\mbox{E-node}\xspace}
\newcommand{\Enodes}{\mbox{E-nodes}\xspace}
\newcommand{\enode}{\mbox{e-node}\xspace}
\newcommand{\enodes}{\mbox{e-nodes}\xspace}

\maketitle

\begin{abstract}
State-of-the-art hardware compilers
  for FPGAs
  often fail to find efficient mappings
  of high-level designs
  to low-level primitives,
  especially complex programmable primitives
  like digital signal processors (DSPs).
\hl{why is approaches plural?}
New approaches apply 
  \textit{sketch-guided program synthesis}
  to more optimally map designs.
However, this approach
  has two primary drawbacks.
First, sketch-guided program synthesis
  requires the user to provide \textit{sketches,}
  which are challenging to write
  and require domain expertise.
Second, the open-source SMT solvers
  which power
  sketch-guided program synthesis
  struggle with the sorts of operations
  common in hardware---%
  namely multiplication.
In this paper, we address both of these challenges
  using an equality saturation (eqsat)
  framework.
By combining eqsat and
  an existing state-of-the-art
  program-synthesis-based tool,
  we produce
  Churchroad,
  a technology mapper
  which handles larger and more complex designs
  than the program-synthesis-based tool alone,
  while eliminating the need
  for a user to provide sketches.
\hl{VC: BY combining equational
  reasoning and an existing sota
  synthesis tool, Lakeroad,
  we scale synthesis based mapping
  to designs that are both larger,
  and requiring more complex reasoning about
  multiplication. <Idea here is to focus on the
  improvement to a ``synthesis-mapping'' approach,
  rather than the contribution of (what I assume is)
  a fairly new and incomplete techmapping tool.}

\end{abstract}


\section{Introduction}


State of the art FPGA hardware synthesis tools
  such as Xilinx's Vivado
  or the open-source Yosys~\cite{wolf2013yosys}
  often fail
  to fully utilize the features of complex,
  programmable FPGA primitives like DSPs~\cite{lakeroad}.
For example,
  they might fail to pack multiple operations
  (e.g.~a multiply followed by an add)
  onto a single DSP,
  instead using a DSP and additional logic resources.

Our recent tool Lakeroad~\cite{lakeroad}
  uses
  \textit{sketch-guided program synthesis}
  to more fully take advantage of all features
  of programmable primitives.
Sketch-guided program synthesis
  is a technique which uses
  SMT solvers to completely search 
  through a space for a solution---%
  in Lakeroad's case, searching through possible DSP configurations.
This technique allows Lakeroad
  to more thoroughly support
  all features of DSPs.
However, Lakeroad
  suffers from two primary limitations.

\textbf{Problem 1:} \textit{User must provide sketches.}
  Lakeroad requires users to provide
  \textit{sketches}~\cite{sketchPhdThesis}.
These sketches capture
  a rough outline
  of the compiled design
  and guide the SMT solvers in
  filling the ``holes'', i.e., missing parts of the design.
Writing sketches is both tedious and difficult,
  requiring
  knowledge of the sketching
  DSL in addition to
  domain expertise
  of the target being compiled to.

  \textbf{Problem 2:} \textit{Scalability of SMT solvers.}
The open-source SMT solvers
  used to power sketch-guided synthesis
  do not scale 
  due to the inherent inefficiency 
  of the underlying bit-blasting 
  algorithm
  typically used in SMT solvers~\cite{bitblasting_wrong_path}.
This prevents
  Lakeroad from
  compiling larger designs or supporting
  operations like multiplication over large bitwidths.
\hl{it's not just multiplication,
  it's also overall size
  of design that's an issue.
Cite control logic synthesis paper
  from zach and jon,
  cite lakeroad.}

This paper mitigates both problems with a key insight:
  a program-synthesis-based tool
  like Lakeroad
  should be viewed as a specialized, high-powered subroutine
  which should only be called on \textit{portions}
  of the design;
  furthermore,
  to orchestrate calls
  to specialized subroutines like Lakeroad, we can use
  a framework called equality saturation (or eqsat)~\cite{egg2021,egglog,smith2024there}.
Eqsat is a term rewriting technique
  built around a core data structure
  called an \egraph~\cite{nelsonPhD}.

This divide-and-conquer approach based on eqsat and \egraph{}s
  helps mitigate
  both aforementioned problems.
First, by facilitating the application of
  semantics-preserving
  rewrite rules~\cite{eqsat2009, egg2021},
  we can break down
  large designs (e.g.~wide multipliers)
  and compile them piece by piece,
  leading to more tractable problems
  for the SMT solvers underlying Lakeroad.
Second, 
  we can eliminate the need for users
  to provide sketches,
  by instead inferring the structure of the output
  from information inside the \egraph.

We showcase this idea
  in Churchroad, a new, open-source\footnote{
  Churchroad is located at \url{https://github.com/gussmith23/churchroad}.
  }
  technology mapper.
Churchroad is built upon the egglog~\cite{egglog}
  eqsat framework, and calls out to
  Lakeroad as a
  specialized subroutine.
In the rest of this paper,
  we demonstrate with a detailed example
  the core ideas behind Churchroad.

\hl{ideally, if we can fit it, one sentence
  or one paragraph about how
  this technique is more broadly applicable/
  an example of how eqsat is good for hw}
 
\section{Technical Contribution}
\label{sec:overview}

To motivate Churchroad
  and to explain how it works,
  we use a running example.
We first describe our example design,
  and describe
  how it should be optimally
  compiled
  onto a Xilinx UltraScale+ FPGA.
We then 
  explain how a single query to Lakeroad
  will fail
  to compile this design.
Finally, we explain
  how Churchroad
  uses eqsat and \egraph{}s
  to break the mapping task
  into multiple simpler queries to Lakeroad.

Consider a multiplier
  which multiplies a 16-bit number $a$
  with a 32-bit number $b$
  and produces the lower 32 bits
  of the result
  (all unsigned):
  
\begin{minted}
[fontsize=\footnotesize]
{systemverilog}
module mul(input [15:0] a, input [31:0] b, 
           output [31:0] o);
  assign o = a * b;
endmodule
\end{minted}

\noindent
We refer to this high-level,
  pre-compilation implementation
  of the design
  as the \textit{specification} or \textit{spec}.
From this point on, we will
  refer to the spec with math
  instead of Verilog.
\hl{Haobin: Maybe start with the math form and show the Verilog code later instead?}
For notational convenience, 
  we split $b$ into two 16-bit halves,
  $b_1$ and $b_0$, where $b = b_1 \mdoubleplus b_0$,
  i.e., the two halves concatenated.
Thus, our spec is:
\begin{equation}
\label{eqn:spec}
    \textsc{Spec}\left(a, b\right) := a \times b \ \left(\text{or}\ a \times \left(b_1 \mdoubleplus b_0\right)\right)
\end{equation}

To map this design onto a Xilinx UltraScale+ FPGA,
  it would be ideal to use the UltraScale+'s
  specialized DSP48E2 primitive,
  which efficiently implements
  multiplication.
Implementing our design requires
  two DSP48E2s,
  arranged in the following way:

\begin{center}
\includegraphics[width=.6\linewidth]{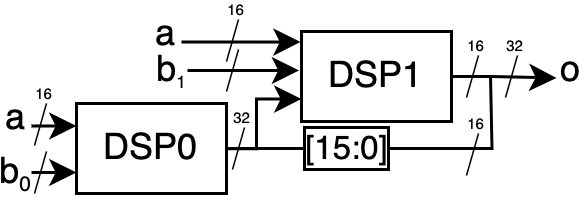}
\end{center}

\noindent
where DSP0 computes
  the lower 16 bits
  of the multiplication,
  $\left(a \times b_0\right)^{15}_0$,
  and
  DSP1 implements
  the upper 16 bits 
  of the multiplication,
  $\left(a \times b_1  + (a \times b_0)\gg 16\right)^{15}_0$
  (extraction of the lower 16 bits is implicit in the diagram),
  taking advantage of
  the DSP's internal shifter\footnote{
  In reality the DSP48E2 uses a 17-bit internal shift;
    we use 16 simply for 
    rounder numbers
    in our example.
  } and the partial product
  produced by DSP0.






\hl{@ chandra: this is the story we should bubble up into the intro, so that this sentence is just a refresher on it.}
Our previous work~\cite{lakeroad}
  demonstrated that existing
  open-source, 
  pattern-matching-based technology
  mappers
  such as those in
  Yosys~\cite{wolf2013yosys}
  struggle to utilize all features
  of complex, programmable primitives
  like DSPs.
In response, we introduced Lakeroad,
  an open-source
  technology mapper
  which can more completely use
  the features of DSPs.
However, Lakeroad was only evaluated
  on small designs
  which map to single DSPs;
  in fact, Lakeroad struggles
  with designs requiring multiple DSPs.
We will now explain exactly how Lakeroad
  will struggle to compile this design,
  to understand 
  the core challenges
  which Churchroad solves.
\looseness=-1

At the core of Lakeroad's
  greater completeness, correctness, and extensibility
  is a technique
  called
  \textit{sketch-guided program synthesis,}
  which depends on user-written \textit{sketches.}
A sketch is an outline
  of the compiled result,
  with \textit{holes}
  which Lakeroad must fill in
  to produce a completed design.
To compile our example design,
  we need a sketch
  which captures the general structure
  of our 2-DSP output.
This sketch currently does not exist within Lakeroad, however,
  which is \textbf{problem 1}:
  users may need to provide their own sketches,
  which can be tricky to write
  and require domain knowledge
  of what the compiled result should look like.
  \looseness=-1


Once written, the sketch looks like\footnote{
We show the sketch here in Verilog for clarity,
  but the sketch is actually written in Lakeroad's Racket-/Rosette-based
  sketching DSL.
}:
\begin{minted}
[fontsize=\footnotesize]
{systemverilog}
module two_dsp_sketch(
    input [15:0] a, input [31:0] b, output [31:0] o);
  logic [31:0] dsp0_out, dsp1_out;
  DSP #(<p0>) DSP0 (.a(a), .b(b[15:0]), .o(dsp0_out));
  DSP #(<p1>) DSP1 (.a(a), .b(b[31:16]), .c(dsp0_out),
                    .o(dsp1_out));
  assign o = {dsp1_out[15:0], dsp0_out[15:0]};
endmodule
\end{minted}

\noindent
It captures
  the structure of the compiled output,
  but leaves the parameters\footnote{
Lakeroad's sketches include holes
  for many of the ports of the DSP
  as well as the parameters,
  but to simplify this example,
  we will simply consider the parameters.
  } of each DSP
  as holes 
  \texttt{<p0>}
  and
  \texttt{<p1>},
  to be filled in by a later step
  of program synthesis.
In mathematical notation, our sketch looks like:
\begin{align}
\label{eqn:sketch}
\begin{split}
\textsc{Sketch}&\left(p_0, p_1, a, b\right) :=\\
\textbf{let}\ &b_1, b_0 := b^{31}_{16}, b^{15}_0 \\
&\text{DSP0} :=\text{DSP}\left(p_0, a, b_0\right) \\
&\text{DSP1} :=\text{DSP}\left(p_1, a, b_1, \text{DSP0}\right) \\
\textbf{in} \ &\text{DSP1}^{15}_0 \mdoubleplus \text{DSP0}^{15}_0
\end{split}
\end{align}

Lakeroad then fills in the holes of the sketch,
  such that the output implements the spec.
Specifically,
  Lakeroad solves
  the following query:
\begin{equation}
\label{eqn:synth-query}
\exists p_0, p_1.\ \forall a, b.\ \textsc{Spec}\left(a,b\right) =\textsc{Sketch}\left(p_0, p_1, a, b\right)
\end{equation}
That is, it searches for a configuration
  of the two DSPs
  (captured by $p_0$ and $p_1$)
  such that $\text{DSP1}^{15}_0 \mdoubleplus \text{DSP0}^{15}_0$ 
  is equivalent to our spec,
  $a\times b$.

To find such a configuration,
  Lakeroad's sketch guided synthesis algorithm
  (implemented by Rosette~\cite{torlak2013growing})
  first
  \textit{guesses}
  at values of
  $p_0$ and $p_1$,
  and then uses an SMT solver
  to verify whether the guesses
  are correct.
To do so, 
  it queries
  an SMT solver, asking
$\forall a, b.\ 
  \textsc{Spec}\left(a, b\right)=
  \textsc{Sketch}\left(p_0, p_1, a, b\right)$
  for the guessed values of 
  $p_0$ and $p_1$.
  
Continuing our example,
  imagine that the program synthesis
  algorithm
  chooses a $p_0$ and $p_1$ such that 
$\forall a, b.\ 
  \textsc{Spec}\left(a, b\right)=
  \textsc{Sketch}\left(p_0, p_1, a, b\right)$
  expands to
\begin{equation}
\label{eqn:simplified-synth-query}
\forall a, b.\ a\times b=
  \left(a \times b_1  + (a \times b_0)\gg 16\right)^{15}_0 \mdoubleplus
  \left(a \times b_0\right)^{15}_0
\end{equation}
If the SMT solver
  can prove this to be true,
  then $p_0$ and $p_1$
  can be used to configure our sketches
  and produce an output design.
If the SMT solver instead returns
  a counterexample---%
  that is, values of $a$ and $b$
  for which \cref{eqn:simplified-synth-query} is not true---%
  it can use the counterexample
  to pick new guesses
  for $p_0$ and $p_1$.

However,
  when we attempt to prove \cref{eqn:simplified-synth-query}
  with
  open-source SMT solvers,
  they
  take an infeasible amount of time
  proving this query---%
  \textbf{problem 2} in the introduction.
To demonstrate this, we implement
  the query directly within the
  SMT solver-aided programming language,
  Rosette~\cite{torlak2013growing}:

\begin{figure}
    \centering
    \includegraphics[width=.99\linewidth]{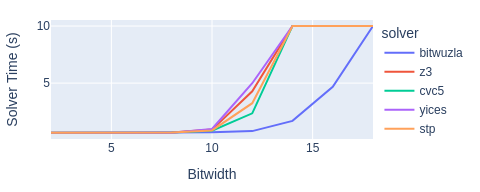}
    \caption{Solver time to prove \cref{eqn:simplified-synth-query} with Rosette, at various settings of \texttt{bw} and with various solvers. Timeout is set to 10 seconds.}
    \label{fig:timeout}
\end{figure}

\clearpage

\begin{minted}
%[xleftmargin=\parindent,linenos,fontsize=\footnotesize]
[fontsize=\footnotesize]
{Racket}
#lang rosette
(define bw 16) ; Larger bitwidths begin to time out!
(define hbw (/ bw 2))
(define-symbolic a b1 b0 (bitvector hbw))
(define b (concat b1 b0))
(verify (assert ; Verify the following assertion:
 (bveq          ; The following are equal:
  (bvmul (zero-extend a (bitvector bw)) b) ; (4) LHS
  (concat                                  ; (4) RHS
   (bvadd (bvmul a b1)
          (extract (- hbw 1) 0
           (bvlshr 
            (bvmul (zero-extend a (bitvector bw))
                   (zero-extend b0 (bitvector bw)))
            (bv hbw bw))))
   (bvmul a b0)))))
 
\end{minted}

\noindent
By adjusting \texttt{bw},
  we observe how state-of-the-art
  SMT solvers behave
  at different bitwidths.
\Cref{fig:timeout}
  shows our results:
  while the underlying solvers are able
  to solve the query
  for small bitwidths,
  around 12 bits,
  most solvers very suddenly
  hit a wall.
At the root of the problem is multiplication.
The underlying bitblasting~\cite{bitblasting_algo}
  algorithm converts the query into 
  first-order logic using
  low-level $\vee$/$\wedge$/$\neg$ operators.
This conversion causes slowdowns
  when faced with bitwise-complex operations like 
  multiplication~\cite{bitblasting_wrong_path}.
  
\textbf{Churchroad's first
  fundamental insight}
  is to mitigate SMT timeouts
  by breaking up 
  large queries
  into multiple smaller queries that
  are easier for SMT solvers to handle.
For example, if we can break
  \cref{eqn:synth-query}
  into two queries,
\begin{align*}
\begin{split}
&\exists p_0.\ \forall a,b.\ a \times b_0 = \text{DSP}(p_0, a, b)\ \text{and} \\
&\exists p_1.\ \forall a,b,c.\ a \times b_1 + c \gg 16 = \text{DSP}(p_1, a, b, c)
\end{split}
\end{align*}
then, after the synthesis procedure
  picks concrete guesses for $p_0$ and $p_1$,
  we might end up with queries like
\begin{align*}
\begin{split}
&\forall a,b.\ a \times b_0 = a \times b_0 \ \text{and} \\
&\forall a,b,c.\ a \times b_1 + c \gg 16 =a \times b_1 + c \gg 16 
\end{split}
\end{align*}
which are trivial for
  an SMT solver to prove.
  
To break up queries, 
  Churchroad pre-applies equalities
  we know to be true.
In this example,
  we'd like to apply
  the equality
  in \cref{eqn:simplified-synth-query}
  to rewrite the spec $a \times b$
  into its expanded form,
  with the eventual goal of compiling its subexpressions.
This equality is not just specific
  to our example---%
  it is true in general,
  and is in Churchroad's database of equalities.
To apply these equalities,
  Churchroad
  utilizes
  the egglog eqsat framework~\cite{egglog},
  which provides
  a data structure
  called an \textit{\egraph}
  to hold our design.
We can then capture
  equalities like \cref{eqn:simplified-synth-query}
  as \textit{rules}
  over the \egraph:
\begin{minted}[fontsize=\footnotesize]{text}
(rule 
 ((= e (Mul arg0 arg1))
  (= (bitwidth arg0) 16) (= (bitwidth arg1) 32))
 ((union e 
   (Concat 
    (Extract 15 0 
     (Add (Mul arg0 (Extract 31 16 arg1))
          (Shr (Mul arg0 (Extract 15 0 arg1)) 16)))
    (Extract 15 0 (Mul arg0 (Extract 15 0 arg1)))))))
\end{minted}

\noindent
egglog rules takes two arguments:
  first, a list of expressions to search for
  in the \egraph,
  and second, a list of commands to run
  if those expressions are found.
In this case,
  this rule looks for 
  a multiplication expression,
  which it names \texttt{e},
  and checks the bitwidths of the arguments
  \texttt{arg0}
  and \texttt{arg1}.
For this example, we have hardcoded this rule
  to search for specific bitwidths, 
  but the actual rule in Churchroad works for variable bitwidths.
If an expression \texttt{e} matching the pattern is found,
  the rule \textit{unions}
  it with the expression
  in the
  right-hand side of \cref{eqn:simplified-synth-query}.
To understand what we mean by \textit{union},
  let's see what this rewrite does
  to our spec.
When we apply this rule to our
  spec (left),
  it matches on the \texttt{Mul} node,
  and produces
  the \egraph on the right:
\begin{center}
\includegraphics[width=.7\linewidth]{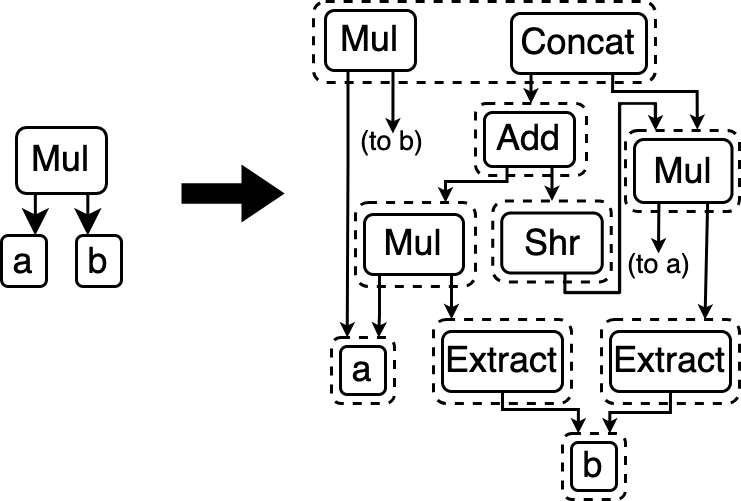}
\end{center}
\noindent
We have simplified the \egraph
  by dropping some extracts.
Note the dotted box
  around the top-level
  \texttt{Mul} and \texttt{Concat} nodes.
This indicates that,
  as a result of the \texttt{union} command in 
  our rule above,
  these nodes
  are in the same \textit{equivalence class}
  or \textit{eclass} within the \egraph,
  and are thus equivalent.
The dotted boxes around the other nodes
  indicate that they are the only nodes
  in their eclasses.

When the \egraph contained only
  \texttt{(Mul a b)},
  the only mapping we could attempt
  was our ill-fated 2-DSP mapping.
Furthermore,
  attempting this mapping required explicit user
  input:
  the user needed
  to provide the sketch
  in \cref{eqn:sketch}.
These sketches are challenging
  to write,
  and require the user
  to have some idea of what the output design
  should look like
  (\textbf{problem 1}).
\textbf{Churchroad's second fundamental insight}
  is that we can 
  entirely remove the need for user-provided sketches.
Instead of requiring the user
  to provide a sketch of the structure of the output,
  we can instead infer the structure of the output
  from information already within the \egraph.
  
To infer the structure of the compiled output,
  Churchroad searches through the \egraph
  for expressions that look like
  they can be implemented using a DSP.
For this, Churchroad   
  again uses egglog rules:
\begin{minted}[fontsize=\footnotesize]{text}
(rule 
 ((= e (Mul arg0 arg1))
  (<= (bw e) 48) (<= (bw arg0) 17) (<= (bw arg1) 17))
 ((union e (DSP? arg0 arg1))))
(rule 
 ((= e (Add (Mul arg0 arg1) (Shl arg2 const)))
  (<= (bw e) 48) ... (<= (bw arg2) 36))
 ((union e (DSP? arg0 arg1 arg2))))
\end{minted}
\hl{HAOBIN: Do we need to explain why we have 48 and 17 here?}

\noindent
These rules search for expressions
  which may be implementable using a DSP.
The first rule simply searches for multiplies;
  the second, for multiply--adds with a shift
  on the input.
Any time these rules fire, they insert a
  DSP proposal node, ``DSP?''.
These nodes simply mark that the eclass
  \textit{may} be implementable using a DSP,
  and that Lakeroad should be used
  to determine whether such a mapping exists.
These rules
  can fire multiple times on the \egraph above,
  resulting in the following \egraph:

\begin{center}
\includegraphics[width=.5\linewidth]{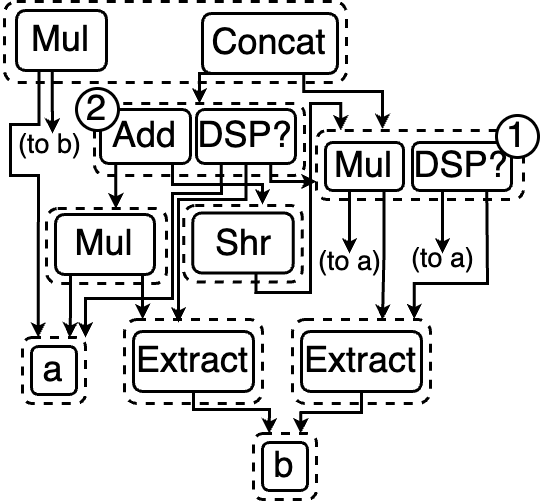}
\end{center}

\noindent
Each eclass with a DSP?~node
  has been identified by our
  rules above
  as potentially implementable
  using a DSP.
To confirm whether or not
  each eclass can actually be implemented
  with a DSP,
  Churchroad now calls Lakeroad as a subroutine
  on each proposed DSP?~node.

Let's begin with the eclass labeled 
  1.
Recall that Lakeroad requires
  both a spec
  and a sketch,
  both of which were provided by the user.
However, Churchroad is able to generate
  both specs and sketches
  automatically, from the information
  in the \egraph.
This eclass contains two nodes:
  one node representing a subexpression
  of the rewritten spec,
  $a \times b_0$,
  and 
  $\text{DSP?}\left(a, b_0\right)$
  representing a potential 
  DSP implementation
  of this eclass.
The first node serves as our spec,
  while the second serves
  as our sketch!
Thus, the synthesis query sent to Lakeroad
  looks like
\begin{equation*}
    \exists p.\ \forall a,b.\ 
    a \times b_0 
  =\text{DSP}\left(p, a, b_0\right)
\end{equation*}
where the sketch 
  $\text{DSP}\left(p, a, b_0\right)$
  was generated by adding parameters $p$
  to the original ``DSP?''  node.
Note that this is much simpler
  than the query in \cref{eqn:synth-query};
  consequently, the underlying SMT solver
  is able to solve it,
  and Lakeroad is able to generate
  a configuration of the 
  UltraScale+ DSP48E2
  which implements
  $a \times b_0$,
  which is then inserted into the 
  \egraph.

We can repeat this process
  for the eclass labeled 2,
  which produces the following synthesis query:
\begin{equation*}
\exists p.\ \forall a,b.\ a \times b_1  + (a \times b_0)\gg 16 
  =
    \text{DSP}\left(p, a, b_1, a \times b_0\right)
\end{equation*}

\noindent
Again, Lakeroad
  returns a DSP mapping
  for this query, which we can insert
  into the \egraph.

The final step
  is to extract a compiled design
  from the \egraph.
Currently, Churchroad
  does this in a straightforward manner:
  for each eclass, it simply chooses
  a node which is legal
  in structural Verilog.
For example, module instantiations,
  zero-/sign-extensions,
  extractions, and concatenations are allowed,
  while behavioral operators like
  Mul are not.
In this case, we extract
  our two DSP instantiations produced by Lakeroad.
Finally, Churchroad converts
  from the internal representation
  extracted from the \egraph
  to structural Verilog,
  producing the design
  in the block diagram 
  at the beginning of the section.
From beginning to end, Churchroad
  currently takes about 4 seconds
  to compile this example.

With that, Churchroad has successfully
  compiled a design
  which Lakeroad could not handle.
Using eqsat and encoding
  and certain equalities
  like \cref{eqn:simplified-synth-query}
  as axioms,
  Churchroad expanded the initial spec.
Then,
  using DSP proposal rules,
  Churchroad found locations in the expanded spec
  which might be implementable using a DSP.
Finally,
  these DSP proposals were sent to Lakeroad,
  which produced mappings to DSPs.
Finally, with all of the Lakeroad-produced mappings
  stored in the \egraph,
  Churchroad extracted the final compiled design.









\section{Related Work}
\label{sec:related-work}

\hl{
certainly cite haploid here: https://github.com/IanBriggs/haploid
}

\hl{more new work on making multiplication scale: https://arxiv.org/pdf/2406.04696}


Programming language and formal methods techniques
  have been used for
  verifying hardware designs~\cite{kami, koika, bluespec}.
Sisco et al.~\cite{zachsisco} used sketch-guided program
  synthesis for ``loop rerolling''---this enabled
  decompilation of low-level netlist to generate high-level HDL.
Other papers have also explored the use of sketch-guided
  synthesis for generating HDL implementations
  from high-level designs~\cite{verisketch, sketchilog}.
Tools like ODIN~\cite{odin} and ODIN-II~\cite{odin2}
  rely on syntactic matching to
  map portions of a design to specialized units.

Brain et al.~\cite{bitblasting_wrong_path}
  demonstrated that bit-blasting can be optimized by changing the multiplier encoding using either; (i) constant propagation and rewrites; or (ii) polynomial interpolation.
Rath et al.~\cite{polysat} have presented an alternative to bit-blasting with PolySAT by providing a theory-solver for bit-vector arithmetic using non-linear polynomials and a bit-vector plugin to the equality graph.
Yu et al.~\cite{formal_verification_function_extraction}'s work enables formal verification of large arithmetic circuits.
There is also a functional verification  technique for multipliers presented by Yu et al.~\cite{verification_galois_field} that uses Galois field (GF) arithmetic to verify an n-bit GF multiplier in n threads, with experimental results up to 571 bits.
Dafny supports user-defined axioms to prevent SMT solver timeouts~\cite{dafny-ref}.
Haploid~\cite{haploid} is another tool that
  uses axioms to simplify SMT queries in a preprocessing step using
  equality saturation.


\hl{
jin: bring your own axioms to SMT solvers? in Dafny.

citations for splitting up a proof to make the proof go through. do people do this in Dafny?
}

\hl{
Functional netlists: https://dl.acm.org/doi/abs/10.1145/1411204.1411253
}

\hl{
jin:
AIR dialect from AMD in MLIR:
https://xilinx.github.io/mlir-air/index.html
}

\hl{
zsisco: there's some related work I wanted to mention here:
 There's a field of research about arithmetic circuit extraction:
 Basically, given a circuit, find all of the multipliers.
 This has applications for formally verifying circuits,
 and experiences similar SAT/SMT blow-ups you experienced.
 There's a relevant line of work I wanted to bring up
 which uses an algebraic approach
 to represent circuits as polynomials in a finite field (GF2),
 and rewrite them to find multipliers (not eq-sat).
 Happy to add a blurb and citations about this if you have a related work section.
 Relevant links:
 \texttt{https://ieeexplore.ieee.org/abstract/document/7442835},
 \texttt{https://ieeexplore.ieee.org/abstract/document/7858326}
}

\section{Conclusions and Future Work}
\label{sec:conclusion}

This work provides early evidence that
  equational reasoning using \egraphs
  can help scale program synthesis-based
  technology mapping techniques.
We demonstrated this idea
  in the form of a new tool, Churchroad,
  which uses Lakeroad as one specialized subroutine
  that focuses on specific portions of the design.
Using equational reasoning,
  Churchroad breaks down larger designs,
  and thus shrinks the queries sent to Lakeroad
  while eliminating the need for user-provided sketches.

In the future,
  we are eager to demonstrate how \egraphs
  can be used to orchestrate more tools like
  Yosys, Lakeroad, ABC, etc. to further improve technology mapping.
We also plan to generate correct rewrite rules
  for Churchroad, rather than writing them manually.



\section*{Acknowledgements}
We thank Zachary D.~Sisco, Vishal Canumalla, and Jin Yang
  for feedback on our work and
  Thanawat Techaumnuaiwit
  for helping with the implementation.

\bibliographystyle{IEEEtran}
\bibliography{references}

\end{document}